\documentclass{aastex}
\usepackage{spr-astr-addons}
\usepackage{url}
\usepackage{graphicx}
\usepackage{amssymb}
\usepackage{amsmath}
\usepackage{stmaryrd}

\RequirePackage{color}

\newcommand{\corot}{{\textsc{CoRoT}}}
\newcommand{\cible}{{HD~49385}}

\newcommand{\ind}[1]{_{\rm #1}}

\def\m2s2{\,m$^{2}$\,s$^{-2}$} %m2.s -2
\def\aov{\alpha_{\hbox{\rm\small ov}}}

\begin{document}

\title{New insights on the interior of solar-like pulsators \\ thanks to CoRoT: the case of HD 49385}
%% Running heads
\shorttitle{}
\shortauthors{Deheuvels et al.}

\author{S. Deheuvels} \and \author{E. Michel}
\affil{LESIA, CNRS UMR 8109,  Observatoire de Paris, Universit\'e Paris 6, Universit\'e Paris 7, 92195 Meudon cedex, France}
\email{sebastien.deheuvels@obspm.fr}

%\altaffiltext{1}{LESIA, CNRS UMR 8109,  Observatoire de Paris, Universit\'e Paris 6, Universit\'e Paris 7, 92195 Meudon cedex, France}

\begin{abstract}

The high performance photometric data obtained with space mission CoRoT
offer the opportunity to efficiently constrain our models for the stellar
interior of solar-like pulsating stars. 
On the occasion of the analysis of the oscillations of solar-like pulsator HD~49385,
a G0-type star in an advanced stage of evolution, we revisit the phenomenon of the avoided
crossings. \cite{coursCD} proposed a simple analogy to describe an avoided crossing between two modes. We here present an extension of this analogy
to the case of $n$ modes, and show that it should lead, in certain cases, to a characteristic behavior of the eigenfrequencies, significantly different
from the $n=2$ case. This type of behavior seems to be observed in HD~49385, from which we infer that the star should be in a Post Main Sequence phase.

\end{abstract}

\keywords{Stellar evolution -- Stellar pulsation -- Stellar interior -- Advanced stages of stellar evolution }

%\section*{}
%\label{sec:intro}

\section{Introduction}

As a star evolves, the eigenfrequencies of the $g$ modes increase due to an increase of the buoyancy frequency. When the frequency of a non-radial $g$ mode becomes close to 
that of a $p$ mode of same degree $\ell$, the two modes undergo an avoided crossing, at the end
of which they have exchanged natures. During the avoided crossing, both modes present a mixed character: they act as $g$ modes in the deep interior, and as $p$ 
modes below the surface (see \textit{e.g.} \citealp{1974A&A....36..107S}, \citealp{1977A&A....58...41A}). 

The avoided crossings happen on very short time-scales compared to the evolution timescale. Besides, the existence of mixed modes and
their frequencies if they exist, depend greatly on the profile of the buoyancy frequency throughout the star. The observation of such modes would therefore
yield strong constraints on the age of the star, and on its inner structure, as described in \cite{1991A&A...248L..11D}. For solar-like pulsators, though models showed 
that mixed modes should be present in the spectrum of certain evolved objects, \textit{e.g.} $\eta$ Bootis \citep{2003A&A...404..341D}, they were never clearly
identified.

Avoided crossings are usually assumed to involve two modes only. It is the case, for exemple, in \cite{coursCD}, where a simple analogy is proposed
to describe the phenomenon. We present in Sect. \ref{sect_analogy}, an extension of this analogy to the case of an avoided
crossing between $n$ modes and apply our results to study the evolutionary status of solar-like pulsator \cible\ in Sect. \ref{sect_model}.

\section{Avoided crossing with $n$ modes \label{sect_analogy}}

\subsection{A simple analogy for $n=2$}

The fact that modes with close eigenfrequencies have a mixed character is due to the coupling which exists between the $g$-mode cavity and the $p$-mode cavity.
If they were uncoupled, two modes of different nature could share the same eigenfrequency, without any impact on their eigenfunctions. Based on an original idea of \cite{1929ZhPhy..30..467V}, 
\cite{coursCD} proposed a simple 
analogy to understand the process of the avoided crossings. He considered the two cavities of the star as a system of two coupled oscillators $y_1(t)$ and $y_2(t)$ with
a time dependence, responding to the following system of equations:
\begin{eqnarray}
\frac{\hbox{d}^2y_1(t)}{\hbox{d}t^2} & = & -\omega_1(\lambda)^2y_1+\alpha_{1,2} y_2 \label{eqdiff_1} \\ 
\frac{\hbox{d}^2y_2(t)}{\hbox{d}t^2} & = & -\omega_2(\lambda)^2y_2+\alpha_{1,2} y_1 \label{eqdiff_2}
\end{eqnarray}
where $\alpha_{1,2}$ is the coupling term between the two oscillators. $\omega_1(\lambda)$ and $\omega_2(\lambda)$ are the eigenfrequencies of the uncoupled
oscillators ($\alpha_{1,2}=0$). They depend on a parameter $\lambda$, used to model the change of the dimensions of the cavities as the star evolves. We suppose
that for a certain $\lambda=\lambda_0$, the uncoupled oscillators cross, \textit{i.e.}
\begin{equation}
\omega_1(\lambda_0)=\omega_2(\lambda_0)\equiv\omega_0
\end{equation}
We aim at determining the eigenfrequency 
$\omega(\lambda)$ of the whole system. The solution of Eq. \ref{eqdiff_1} and \ref{eqdiff_2} is under the form:
\begin{eqnarray}
y_1(t)=c_1 \exp(-i\omega t) \label{eq_1} \\
y_2(t)=c_2 \exp(-i\omega t) \label{eq_2}
\end{eqnarray}
By inserting Eq. \ref{eq_1} and \ref{eq_2} into Eq. \ref{eqdiff_1} and \ref{eqdiff_2}, we find that the eigenfrequencies of the system are obtained by solving the eigenvalue
problem $AC=\omega^2 C$, where
\begin{equation}
A=
\begin{bmatrix}
\omega_1^2 & -\alpha_{1,2} \\
-\alpha_{1,2} & \omega_2^2
\end{bmatrix}
\end{equation}
and $C=[c_1, c_2]$.

We get the two following solutions for the system:
\begin{equation}
\omega_{\pm}^2=\frac{\omega_1^2+\omega_2^2}{2}\pm\frac{1}{2} \left[ (\omega_1^2-\omega_2^2)^2+4\alpha_{1,2}^2 \right] ^{1/2} \label{sol12}
\end{equation}
We verify here that if the coupling term $\alpha_{1,2}$ is very small compared to the difference between the eigenfrequencies ($\alpha_{1,2}\ll |\omega_1^2-\omega_2^2|$),
then the eigenfrequencies of the system are close to $\omega_1$ and $\omega_2$. If, on the contrary, $|\omega_1^2-\omega_2^2|\ll \alpha_{1,2}$,
then the eigenfrequencies can be approximated by $\omega_{\pm}^2=\omega_0^2\pm \alpha_{1,2}$. We see here that the two oscillators "avoid" the frequency $\omega_0$,
and the larger $\alpha_{1,2}$, the further away of $\omega_0$ they are during the avoided crossing.

\begin{figure}[t]
\includegraphics[width=8cm]{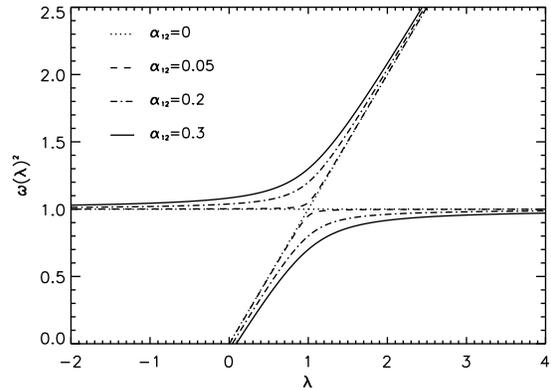}
\caption{Variations of the eigenfrequencies $\omega_{\pm}$ of the system with parameter $\lambda$, for different values of the coupling term $\alpha_{1,2}$.
\label{fig_cross12}}
\end{figure}

\begin{figure}[t]
\includegraphics[width=8cm]{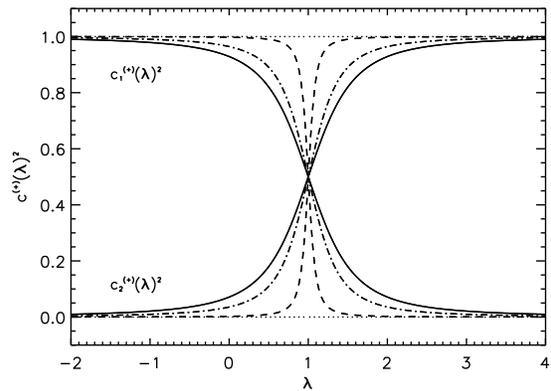}
\caption{Variations of the coefficients $c_1^{(+)}$ and $c_2^{(+)}$ with parameter $\lambda$, for different values of $\alpha_{1,2}$ (the lines correspond to the same values of $\alpha_{1,2}$
as in Fig. \ref{fig_cross12}). We note that the coefficients corresponding to the other eigenfrequency $\omega^{(-)}(\lambda)$ are obtained by interchanging $c_1$ and $c_2$
($c_1^{(-)}(\lambda)=c_2^{(+)}(\lambda)$ and $c_2^{(-)}(\lambda)=c_1^{(+)}(\lambda)$).
\label{fig_cross_coef12}}
\end{figure}

To analyze these solutions in more details, we impose the variations of the eigenfrequencies with $\lambda$: $\omega_1(\lambda)^2=1$ and $\omega_2(\lambda)^2=\lambda$
for example. In this case, we can plot the eigenfrequencies $\omega_{\pm}^2(\lambda)$ (see Fig. \ref{fig_cross12}). Normalizing the coefficients in such a way that
$c_1(\lambda)^2+c_2(\lambda)^2=1$, we obtain the variations of the coefficients with $\lambda$, shown in
Fig. \ref{fig_cross_coef12}. We clearly see that the modes exchange natures during the avoided crossing.

The choice we made for the profile of $\omega_1(\lambda)$ and $\omega_2(\lambda)$ is not an innocent one. Indeed, if we represent the mode frequencies normalized by 
$\sqrt{GM/R^3}$, the $p$-mode frequencies are almost constant during an avoided crossing, while the $g$-mode frequencies increase.

\subsection{Extension of the analogy to the case $n>2$}

The previous analogy relies on the fact that two modes only are affected during an avoided crossing: this corresponds to neglecting the coupling between the two
considered modes, and the other modes in the spectrum. Let us now push the analogy a step further, and consider the case where these other coupling terms 
play a role. We consider $n$ oscillators, instead of two only. We take $n-1$ of them to have constant eigenfrequencies with $\lambda$, $\omega_i^2=i$, $i=1,...,n-1$
(to simulate $p$ modes), and one of them to variate as $\omega_n^2=\lambda$ (to simulate a $g$ mode). We introduce a coupling between the $g$ mode
and all the $p$ modes, resulting in the following system of equations:
\begin{eqnarray}
\label{eqdiff_n}
\frac{\hbox{d}^2y_1(t)}{\hbox{d}t^2} & = & -\omega_1(\lambda)^2y_1+\alpha_{1,n} y_n \\  \nonumber
\vdots & & \\ \nonumber
\frac{\hbox{d}^2y_{n-1}(t)}{\hbox{d}t^2} & = & -\omega_{n-1}(\lambda)^2y_{n-1}+\alpha_{n-1,n} y_n \\ \nonumber
\frac{\hbox{d}^2y_n(t)}{\hbox{d}t^2} & = & -\omega_n(\lambda)^2y_n+\alpha_{1,n} y_1+\hdots+ \alpha_{n-1,n} y_{n-1}\\ \nonumber
\end{eqnarray}
For simplicity, we assume here that the term of coupling with the $g$ mode is equal to the same value $\alpha$ for all the different $p$ modes. 
By writing the different oscillators $y_i(t)=c_i\exp(-i\omega t)$, the eigenfrequencies of the system are once again found by solving the eigenvalue
problem $AC=\omega^2 C$ with
\begin{equation}
A=\begin{bmatrix}
\omega_1^2 & 0 & \cdots & 0 & -\alpha \\
0 & \omega_2^2 & & \vdots & -\alpha \\
\vdots & & \ddots & & \vdots \\
0 & \cdots &  & \omega_{n-1}^2 & -\alpha \\
-\alpha & \cdots & & -\alpha & \omega_n^2
\end{bmatrix}
\end{equation}
and $C=[c_1, \cdots, c_n]$.

\begin{figure}[t]
\includegraphics[width=8cm]{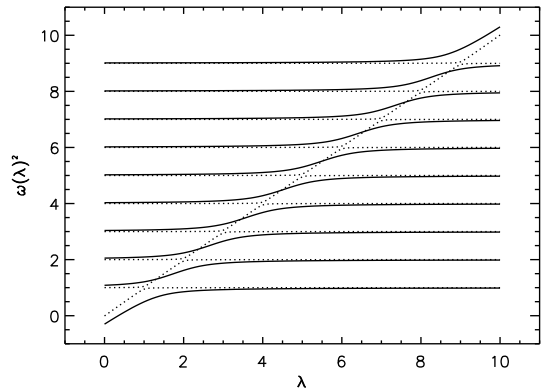}
\caption{Variations of the eigenfrequencies of $(n-1)$ $p$ modes coupled to a $g$ mode which undergoes avoided crossings with the $p$ modes (here, $n=10$).
The dashed lines correspond to a "weak coupling" ($\alpha=0.05$), and the full lines to a "strong coupling" ($\alpha=0.35$).
\label{fig_cross_n}}
\end{figure}

The solution is plotted in Fig. \ref{fig_cross_n} for two different values of $\alpha$.

\begin{figure}[t]
\includegraphics[width=8cm]{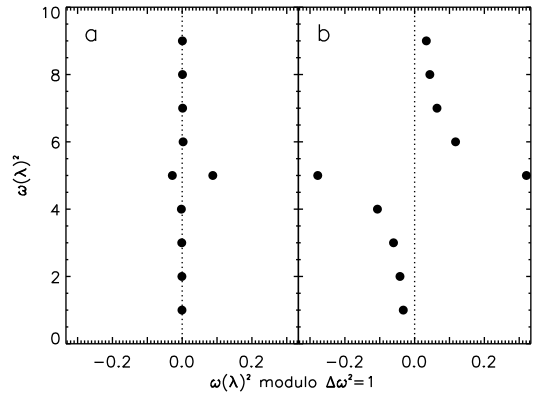}
\caption{The eigenfrequencies for a certain value of $\lambda$ are presented in an \'echelle diagram. In \textit{a}, we observe the perturbation of the ridge due to an 
avoided crossing in the case of a weak coupling ($\alpha=0.05$). In \textit{b}, the perturbation in the case of a strong coupling ($\alpha=0.35$). In both
figures, the dashed line shows the ridge the $n-1$ first oscillators would follow without avoided crossing (or during an avoided crossing with no coupling between the
modes).
\label{fig_ech_n}}
\end{figure}

Once again, the choice of the evolution of the $\omega_i$ with $\lambda$ was made on purpose. The $n-1$ modes simulating the $p$ modes were taken equidistant,
as are the $p$ modes at the first order of the asymptotic theory. For each value of $\lambda$, we can plot an \'echelle diagram, which is a commonly used tool
to characterize equidistances. In the case of uncoupled oscillators, we obtain a straight ridge (see Fig. \ref{fig_ech_n}). 
Fig. \ref{fig_ech_n} shows the impact of the avoided crossings on the curvature of the ridge: in the case of a weak coupling between the
modes, the approximation that only two modes are affected is legitimate since the rest of the ridge is the same as for $\alpha=0$. However, as the coupling
term increases, this ceases to be true, and we can observe a significative change in the curvature of the ridge.

\subsection{Interest in stellar modelling}

Our analogy suggests that, provided the coupling between the $p$-mode cavity and the $g$-mode cavity is strong enough, an avoided crossing does not affect only
two modes of same degree $\ell$, but also a certain number of neighboring modes of degree $\ell$ which act mainly as $p$ modes, but also have a small $g$-mode behavior in the center.
This results in a distortion of the ridge of degree $\ell$ such as that shown in Fig. \ref{fig_ech_n}b. The existence of such a distortion in the spectrum
of an evolved star would make it much easier to detect the presence of a mixed mode and could give constraints on the inner structure of the star, as will be stressed in 
Sect. \ref{sect_perspectives}.

The strength of the coupling between the two cavities is inversely proportional to the size of the evanescent zone which separates them.
The coupling will therefore be strongest for modes of low degree $\ell$, since their Lamb frequency $S_{\ell}$ is smaller. We therefore expect
the distortion of the ridge to be most significant for modes of low $\ell$. This is interesting since
the modes of low degree are those we are most likely to observe in stars. We also remark that, since avoided crossings occur later in the
evolution of the star for modes of lower $\ell$, we expect to observe this distortion only for very evolved objects.

\section{Application to \cible \label{sect_model}}

\cible\ is a solar-like pulsator, of type G0 and was observed with space mission \corot\ during 137 days.
Through a detailed analysis, \cite{analyse} found for the star $T\ind{eff}=6095 \pm50$ K, $\log g=4.00 \pm 0.05$ and a metallicity of
[Fe/H]=$0.09 \pm 0.05$ dex.
The solar-like oscillations of \cible\ were analyzed in \cite{analyse}. 
An extensive modelling of the star is out of scope here, and will be presented in a future work. Our aim here is not to find an optimal
model, but to show that the phenomenon we described in Sect. \ref{sect_analogy} seems to occur in \cible, and that it gives precious information
regarding the evolutionary status of the star.

\subsection{Properties of the model}

Models were computed with the evolution code CESAM2k \citep{1997A&AS..124..597M}, and the mode frequencies were derived from the models 
using the Liege Oscillation Code (LOSC, see \citealp{2008Ap&SS.316..149S}).

We used, for all our models, the OPAL 2001 equation of state, and opacity tables as described in \cite{2008Ap&SS.316..187L}. 
The nuclear rates are computed using the NACRE compilation \citep{1999NuPhA.656....3A}.
Our models use the abundances of \cite{1993oee..conf...15G}. The atmosphere is described by Eddington's grey temperature - optical depth law. 
The convective regions are treated using the Canuto, Goldman, Mazzitelli (CGM) formalism \citep{1996ApJ...473..550C} calibrated on the Sun.

For this preliminary modeling, we considered models not including microscopic diffusion. An arbitrary amount of overshooting (extension of the convective core
over a fraction $\aov=0.1$ of the pressure height scale) was added.

\subsection{Evolutionary stage of \cible}

\begin{figure}[t]
\includegraphics[width=8cm]{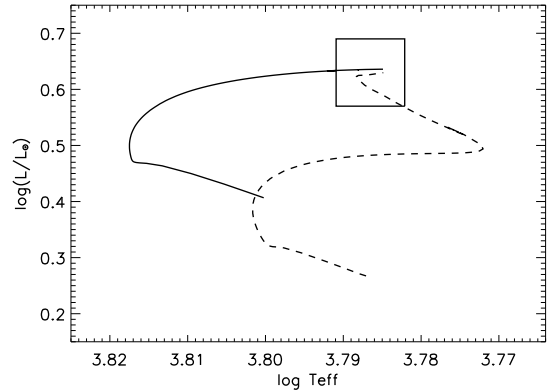}
\caption{Two evolutionary tracks fitting the position of \cible\ in the HR diagram. The stellar parameters of \cible\ are indicated within 1-$\sigma$ error bars by the box.
The full line corresponds to a MS model, and the dashed line, to a PoMS model.
\label{fig_deg}}
\end{figure}

Fig. \ref{fig_deg} shows that we can find models both in the Main Sequence (MS) and Post Main Sequence (PoMS), which fit the position of \cible\ in the HR diagram.
We are therefore facing a problem which is quite classical for evolved object: a degeneracy of MS models and PoMS models which results in an uncertainty on the evolutionary
stage of the studied object (see \textit{e.g.} Procyon A in \citealp{2006A&A...460..759P}, $\eta$ Boo in \citealp{2003A&A...404..341D}).
The use of a mean value of the large spacing does not either provide enough constraint on the star to discriminate between the two stages of evolution.

To investigate further on this problem, one must make use of the fact that MS models and PoMS models fitting the position of \cible\ in the HR diagram have a very different structure in
their deep interior. Indeed, MS models are still burning hydrogen in the core, which is convective, whereas PoMS models are burning hydrogen in shell. Their core is no longer
convective, and almost isothermal due to the absence of nuclear reactions. We also notice that no convective region is attached to the H-burning layer in these models.

\begin{figure}[t]
\includegraphics[width=8cm]{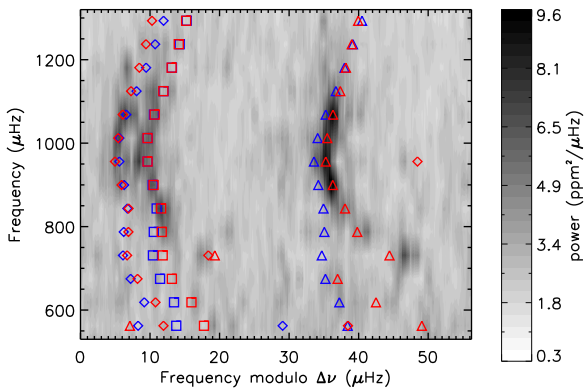}
\caption{\'Echelle diagram of the power spectrum of \cible, computed with a large spacing of $\Delta\nu=56.2\,\mu$Hz. The frequencies of the modes are overplotted
for models \textbf{A} (blue) and \textbf{B} (red). Squares represent $\ell=0$ modes, triangles, $\ell=1$ modes and diamonds, $\ell=2$ modes. A small frequency shift 
(of the order of 1 $\mu$Hz) is applied for both models to partially correct for surface effects.
\label{fig_ech_superpose}}
\end{figure}

Fig. \ref{fig_ech_superpose} shows an \'echelle diagram of the power spectrum of \cible, on which we can see that the curvatures of the $\ell=0$ and $\ell=1$ ridges
are quite different. In fact, the distance between the $\ell=0$ ridge and the $\ell=1$ ridge is sensitive to the inner parts of the star (see \textit{e.g.} \cite{2006A&A...460..759P}). 
A first attempt was made to find models which could reproduce the position of \cible\ in the HR diagram, its large spacing, 
and the distance between the $\ell=0$ and $\ell=1$ ridges at the same time. In the Main Sequence, no such model was found. We plotted in 
Fig. \ref{fig_ech_superpose} an \'echelle diagram of the eigenfrequencies for one of the models fitting the other parameters (further referred to as model \textbf{A}). 
The agreement for $\ell=0$ and $\ell=2$ ridges is satisfactory, but the $\ell=1$ ridge is very poorly reproduced.
On the contrary, we found a PoMS model with the right $T\ind{eff}$, $L$, $\overline{\Delta\nu}$, 
and fitting the $\ell=1$ ridge much better (model \textbf{B}). In this model, we remark the presence of an $\ell=1$ avoided crossing around $\nu=750\,\mu$Hz,
which induces a distortion of the ridge very similar to the one predicted by the analogy we presented in Sect. \ref{sect_analogy}. The observed $\ell=1$ ridge
seems to show the same bend, which cannot be reproduced by MS models since they are not evolved enough to show $\ell=1$ mixed modes
at the appropriate frequency. Under the assumptions we made about the physics of the star, our models are clearly 
in favor of a post Main Sequence status for \cible. This result however needs to be confirmed by further investigation, and a more thorough modeling of the star.

%\begin{figure}[t]
%\includegraphics[width=8cm]{fig_d01_mod.ps}
%\caption{Variations of the $\zeroun$ small spacing. The observations are represented by filled circles, the vertical lines indicating 1-$\sigma$ error-bars. The full line
%corrsponds to a MS model, and the dashed line to a PoMS model.
%\label{fig_d01_mod}}
%\end{figure}

\section{Conclusion and perspectives \label{sect_perspectives}}

We presented the study of an avoided crossing involving $n$ modes,  by extending an analogy developed in \cite{coursCD} to describe 
a two-mode avoided crossing. We showed that for modes of low degree $\ell$ and provided the star is evolved enough, the effect of
avoided crossings should be visible on more than two modes, and the presence of an avoided
crossing should induce a characteristic and easily identifiable distortion of the ridge of degree $\ell$ in the \'echelle diagram, thus
making it easier to detect the presence of mixed modes in the spectrum of the star. We remark that this distortion will be strongest for $\ell=1$ modes, 
and it could possibly be used to identify ridges in pulsating stars evolved enough to show avoided crossings.

As stated before, the frequency at which an avoided crossing occurs in the spectrum of an evolved star varies much faster than the frequencies of the $p$ modes.
This is all the more true for PoMS stars, as seems to be the case for \cible, since in their interior, the increase of the buoyancy frequency $N$
is not only due to the increasing density in the center, but also to the gradient
of chemical composition created by the burning of hydrogen in shell. The observation of an avoided crossing in the $\ell=1$ ridge of \cible\ should therefore lead
to very strong constraints on the age of the star.

Besides, the distortion we observe in the $\ell=1$ ridge is immediately related to the size of the evanescent zone between the cavities. This zone is limited by the
turning point of the $p$-mode cavity, where $\omega=S_{\ell}$, and the turning point of the $g $-mode cavity, where $\omega=N$. 
The location of the latter turning point depends directly on the chemical composition in the center of the star.
By obtaining information on the size of the evanescent zone, we can therefore hope to constrain processes of transport of chemical elements in the inner parts of the
star. For example, even though PoMS models for \cible\ do not have a convective core, they had one during the whole Main Sequence, and we expect the distortion
of the ridge to be sensitive to the amount of overshooting which existed at the boundary of the core.

\acknowledgments
This work was supported by the Centre National d'Etudes Spatiales (CNES). It is based on observations with \corot.

\nocite{*}
\bibliographystyle{spr-mp-nameyear-cnd}
%\bibliography{myref}
\bibliography{biblio-u1}

\begin{thebibliography}{}
\bibitem[Aizenman et al.(1977)]{1977A&A....58...41A} Aizenman, M., Smeyers, P., \& Weigert, A.\ 1977, \aap, 58, 41 
\bibitem[Angulo et al.(1999)]{1999NuPhA.656....3A} Angulo, C., et al.\ 1999, Nuclear Physics A, 656, 3
\bibitem[Canuto et al.(1996)]{1996ApJ...473..550C} Canuto, V.~M., Goldman, I., \& Mazzitelli, I.\ 1996, \apj, 473, 550 
%\bibitem[Christensen-Dalsgaard(1981)]{1981MNRAS.194..229C} Christensen-Dalsgaard, J.\ 1981, \mnras, 194, 229 
\bibitem[Christensen-Dalsgaard (2003)]{coursCD} Christensen-Dalsgaard, J., Lecture Notes on Stellar Oscillations 2003
\bibitem[Deheuvels et al.(2009)]{analyse} Deheuvels, S. et al. 2009, submitted to \aap
\bibitem[Di Mauro et al.(2003)]{2003A&A...404..341D} Di Mauro, M.~P., Christensen-Dalsgaard, J., Kjeldsen, H., Bedding, T.~R., \& Patern{\`o}, L.\ 2003, \aap, 404, 341 
\bibitem[Dziembowski \& Pamyatnykh(1991)]{1991A&A...248L..11D} Dziembowski, W.~A., \& Pamyatnykh, A.~A.\ 1991, \aap, 248, L11 
\bibitem[Grevesse \& Noels(1993)]{1993oee..conf...15G} Grevesse, N., \& Noels, A.\ 1993, Origin and Evolution of the Elements, 15 
\bibitem[Lebreton et al.(2008)]{2008Ap&SS.316..187L} Lebreton, Y., Montalb{\'a}n, J., Christensen-Dalsgaard, J., Roxburgh, I.~W., \& Weiss, A.\ 2008, \apss, 316, 187 
\bibitem[Morel(1997)]{1997A&AS..124..597M} Morel, P.\ 1997, \aaps, 124, 597 
\bibitem[Provost et al.(2006)]{2006A&A...460..759P} Provost, J., Berthomieu, G., Marti{\'c}, M., \& Morel, P.\ 2006, \aap, 460, 759 
\bibitem[Scuflaire(1974)]{1974A&A....36..107S} Scuflaire, R.\ 1974, \aap, 36, 107 
\bibitem[Scuflaire et al.(2008)]{2008Ap&SS.316..149S} Scuflaire, R., Montalb{\'a}n, J., Th{\'e}ado, S., Bourge, P.-O., Miglio, A., Godart, M., Thoul, A., \& Noels, A.\ 2008, \apss, 316, 149
\bibitem[von Neuman \& Wigner(1929)]{1929ZhPhy..30..467V} von Neuman, J., \& Wigner, E.\ 1929, Zhurnal Physik, 30, 467 

\end{thebibliography}

\end{document}